\documentclass[aps,pra,twocolumn,superscriptaddress,floatfix,longbibliography]{revtex4-2}
\usepackage{amsmath}
\usepackage{amssymb}
\usepackage{bm}
\usepackage{graphicx}
\usepackage{color}
\usepackage[normalem]{ulem}
\usepackage{soul}
\usepackage{url}

\usepackage{verbatim}

\usepackage{hyperref}
\hypersetup{pdftex,colorlinks=true,allcolors=blue,bookmarksnumbered=true}
\usepackage{hypcap}

\usepackage{graphicx}
\usepackage{bm}
\usepackage{float}
\usepackage{booktabs}
\usepackage{tabularx}
\usepackage{colortbl}
\usepackage{etoolbox}
\usepackage{subfig}
\usepackage{soul}

\begin{document}


\title{Searching for Low-Mass Axions using Resonant Upconversion}

\author{Catriona A. Thomson}
\email{catriona.thomson@research.uwa.edu.au}
\affiliation{Quantum Technologies and Dark Matter Labs, Department of Physics, University of Western Australia, 35 Stirling Highway, Crawley, WA 6009, Australia.}
\author{Maxim Goryachev}
\affiliation{Quantum Technologies and Dark Matter Labs, Department of Physics, University of Western Australia, 35 Stirling Highway, Crawley, WA 6009, Australia.}
\author{Ben T. McAllister}
\affiliation{Quantum Technologies and Dark Matter Labs, Department of Physics, University of Western Australia, 35 Stirling Highway, Crawley, WA 6009, Australia.}
\affiliation{Centre for Astrophysics and Supercomputing, Swinburne University of Technology, John St, Hawthorn VIC 3122, Australia}
\author{Eugene N. Ivanov}
\affiliation{Quantum Technologies and Dark Matter Labs, Department of Physics, University of Western Australia, 35 Stirling Highway, Crawley, WA 6009, Australia.}
\author{Paul Altin}
\affiliation{ARC Centre of Excellence For Engineered Quantum Systems, The Australian National University, Canberra ACT 2600 Australia}
\author{Michael E. Tobar}
\email{michael.tobar@uwa.edu.au}
\affiliation{Quantum Technologies and Dark Matter Labs, Department of Physics, University of Western Australia, 35 Stirling Highway, Crawley, WA 6009, Australia.}

\date{\today}

\begin{abstract}
We present new results of a room temperature resonant AC haloscope, which searches for axions via photon upconversion. Traditional haloscopes require a strong applied DC magnetic background field surrounding the haloscope cavity resonator, the resonant frequency of which is limited by available bore dimensions. UPLOAD, the UPconversion Low-Noise Oscillator Axion Detection experiment, replaces this DC magnet with a second microwave background resonance within the detector cavity, which upconverts energy from the axion field into the readout mode, accessing axions around the beat frequency of the modes. Furthermore, unlike the DC case, the experiment is sensitive to a newly proposed quantum electromagnetodynamical axion coupling term $g_{aBB}$. Two experimental approaches are outlined - one using frequency metrology, and the other using power detection of a thermal readout mode. The results of the power detection experiment are presented, which allows exclusion of axions of masses between 1.12 $-$ 1.20 $\mu eV$ above a coupling strength of both $g_{a\gamma\gamma}$ and $g_{aBB}$ at $3 \times 10^{-6}$ 1/GeV, after a measurement period of 30 days, which is a three order of magnitude improvement over our previous result.

\end{abstract}

\maketitle

\section{Introduction}

\begin{figure}[t!]
\includegraphics[width=0.60\columnwidth]{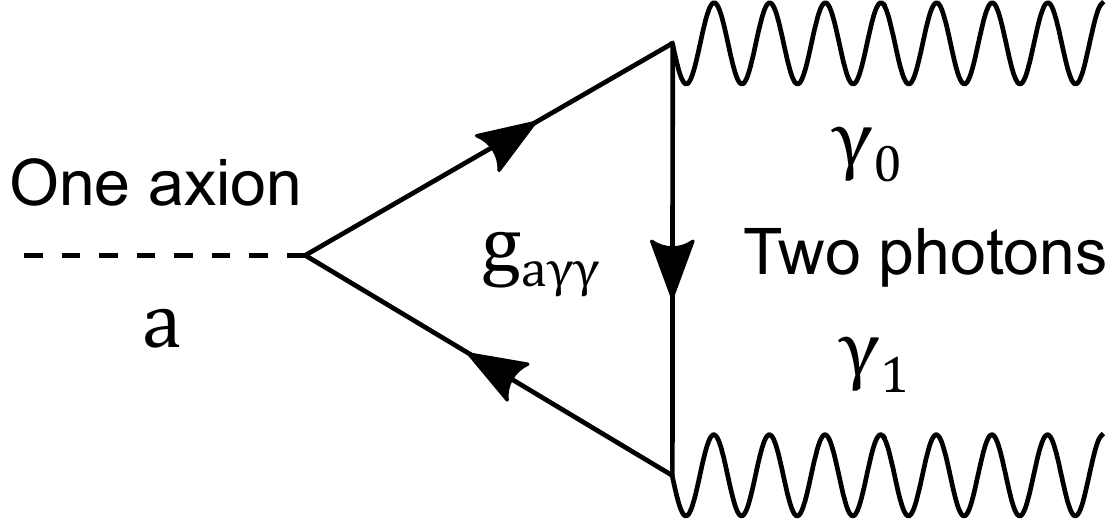}
\caption{Feynman diagram of the axion, $a$, interacting with two photons ($\gamma_0$, $\gamma_1$), via the electromagnetic anomaly of coupling strength $g_{a\gamma\gamma}$.}
\label{2photon}
\end{figure}

The axion is a putative neutral spin-zero boson necessary to solve the strong charge-parity problem in quantum chromodynamics (QCD), which inherently couples weakly to known particles and is predicted to be produced abundantly in the early universe, and is thus a popular candidate for cold dark matter \cite{PQ1977,Wilczek1978,Weinberg1978,wisps,Preskill1983,Sikivie1983,Sikivie1983b}. Axions are commonly searched for via the two-photon electromagnetic anomaly (see Fig.~\ref{2photon} for Feynman diagram) determined in QCD via the axion mixing with the neutral pion. Historically, searches have targeted two well-known models for axion coupling strength, the (Kim–Shifman–Vainshtein–Zakharov) KSVZ \cite{K79,SVZ80,Kim2010} and (Dine–Fischler–Srednicki–Zhitnitsky) DFSZ \cite{DFS81,Zhitnitsky:1980tq,Dine1983} predictions. However, more recently the axion search space has been expanded to include the possibility of axion-like particles being produced in the early universe and the possibility of so-called ``photophilic" and ``photophobic" axions \cite{Svrcek_2006,Arvanitaki10,Higaki_2013,Baumann16,Co2020,Co2020b,Co2021,Oikonomou21,Sikivie2021,Sokolov:2021uv,DILUZIO20201,Rodd2021}. This means the axion mass, $m_a$, could be extremely light or extremely heavy, which motivates searching a wide span of axion parameter space. 

In this work we study the case where an ensemble of axions interacts with an ensemble of photons (the input, pump or background photons at frequency $\omega_0$) to perturb or produce a very weak readout photon signal of frequency  $\omega_1\sim\omega_0\pm\omega_a$, via the two-photon axion anomaly, where $\omega_a\equiv m_a$. Here, we consider only low-mass axions in the quasi-static limit, so that $\omega_a<<\omega_0,\omega_1$, where the interacting fields are essentially classical. Such interactions upconvert the low mass axion signal at $\omega_a$ to near the photonic frequencies of $\omega_0$ and $\omega_1$. UPLOAD (the UPconversion Low-Noise Oscillator Axion Detection experiment) operates on this principle. Furthermore, it has been recently proposed that the upconversion technique is additionally sensitive to a newly theorized axion coupling to photons in quantum electromagnetodynamics (QEMD), $g_{aBB}$, which is predicted to exist if high energy magnetic charge exists \cite{SokolovMonopole22,TobarQEMD22}. A DC (direct current) magnetic field haloscope is not sensitive to this coupling (as derived in \cite{TobarQEMD22}),

This type of low-mass upconversion experiment was first proposed in \cite{Goryachev2019} and experimentally realized in \cite{Cat21}, and showed that two modes excited in the same resonant cavity, with non-zero overlap of the electric field of one with the magnetic field of the other, would cause a putative axion background of dark matter to perturb the frequency (or phase) and amplitude (or power) of the readout mode. The former effect is probed in what we call the ``frequency technique'' and the latter using the ``power technique''. The first prototype experiment of the ``frequency technique'' was performed in \cite{Cat21}, which looked for phase or frequency variations imprinted on the readout oscillator. In 2010, Sikivie put forward the concept of the ``power technique'' \cite{Sikivie2010}, which focused on the search for larger mass axions using downconversion, i.e. configurations where axion signals at higher frequencies were downconverted to lower frequencies. More recently, variations of the ``power technique'' have been proposed, which excite only the background mode, then look for power generated at the readout mode frequency via upconversion \cite{berlin2020axion,Lasenby2020,Lasenby2020b,ABerlin2021}, as is our aim in the reported experiment. 

The DC haloscope remains the most common way to search for axions via power conversion where the background field is a DC magnetic field $(\omega_0=0)$. Historically, the reason for pursuing power detection over frequency detection is because the DC Sikivie halocope \cite{Sikivie83haloscope,Sikivie1985} is only second order sensitive to frequency shifts \cite{tobar2021abraham}. However, when two AC (alternating current) modes are excited, the situation is different, and the DC and AC haloscopes belong to different classes of detectors. Since virtual photons or static fields carry no phase, the DC haloscope belongs to the class of phase-insensitive systems. In contrast, the AC scheme relies on a pump signal carrying relative phases to the readout signal and axion field. This is analogous to existing amplifiers that can be grouped into DC (phase insensitive) amplifiers, where energy is drawn from a static power supply, and parametric (phase sensitive) amplifiers, where energy comes from oscillating fields \cite{Caves82}.

In this work we consider the AC haloscope technique to search for low-mass axions to avoid the large volume DC magnets needed for experiments like SHAFT \cite{Gramolin:2021wm}, Dark Matter Radio \cite{SilvaFeaver2017,Phipps2020} and ABRACADABRA \cite{ABRACADABRA,Ouellet2019FirstRF,Oue19,ABRA21}. Without the use of such large magnetic fields one might think that the technique is not worth pursuing, however, the technique is compensated by a few factors; first the effective Q (quality factor) of the virialized axion is enhanced through the upconversion process, and also, due to the absence of a large magnetic field, high-Q superconducting cavities may be used along with superconducting readout electronics without the problem of magnetic shielding. In the following we compare the frequency and power techniques both theoretically and experimentally, with the goal of optimizing our experiment over the chosen mass range, to below the mass range of the Axion Dark Matter eXperiment (ADMX) \cite{Du2018,Braine2020,Bartram2021} and above the domain of so-called ultra-light axions. We also present the first limits using the upconversion power technique in a prototype room temperature experiment.

\section{Resonant Axion Upconversion}

Axion-photon upconversion experiments require the configuration of two microwave modes in a cavity resonator with a non-zero overlap of respective E (electric) and B (magnetic) fields, where one mode is excited with high power (the pump mode), while the other is used to read out a signal from the axion mixing with the pump mode. There are two types of axion-photon upconversion experiments: 1) the case where both the pump and readout modes are actively excited in the cavity, where the frequency of the readout mode is the physical observable of the measurement (we call this the frequency technique)~\cite{Goryachev2019,Thomson:2021wk,Cat21}; 2) the case where only the pump mode is excited in the cavity, and the thermal/axion-induced amplitude of the quiet readout mode is the physical observable (the power technique)~\cite{berlin2020axion,Lasenby2020b,ABerlin2021,Lasenby2020}. 

In other work we have implemented perturbation theory to calculate the expected frequency shift \cite{TobarQEMD22}, and show it produces a result consistent with our previous derivations for the frequency technique \cite{Goryachev2019,Thomson:2021wk,Cat21}, while to derive the sensitivity for the power technique, complex Poynting theorem has been implemented \cite{TobarQEMD22}. For low-mass axion searches, this technique is limited by how close two modes can be tuned to each other. In particular, one could imagine problems if the modes exist within one another's bandwidth. To access ultra-light axions, this challenge may be overcome by probing just a single mode with non-zero helicity, where the mode acts as its own background field, an idea which is explored in Ref.~\cite{Anyon22}. Furthermore, we have generalized these calculations to include the monopole coupling terms \cite{SokolovMonopole22,TobarQEMD22}, whereby we find upconversion experiments to be additionally sensitive to the axion-photon coupling term, $g_{a BB}$, defined in \cite{SokolovMonopole22}, so in our case limits set on $g_{a\gamma\gamma}$, the standard axion-photon coupling parameter, are equivalent to limits on $g_{a BB}$. Finally, it is worth mentioning that this type of axion haloscope may also be sensitive to high frequency gravitational waves \cite{Berlin2022,sym14102165}.

Axion to photon conversion in the cavity occurs through the dissipative channel associated with the electrical currents in the cavity walls as explained by Poynting theorem \cite{TobarQEMD22,tobar2021abraham}. Similar to an antenna, the oscillating currents will radiate electromagnetic fields, in this case into the cavity volume. The loss takes place at the cavity walls where the magnetic and electric fields are in phase (or current and voltage), while the radiated photons oscillate over the cavity volume out of phase. On resonance, the reactance of the electric and magnetic fields cancels, but a high Q-factor allows power to build up, making the experiment's sensitivity proportional to the cavity Q-factor.

\subsection{Sensitivity of Axion Upconversion Experiments}

Here we derive the sensitivity of the power technique, where the background field will mix with the axion to generate power at the readout mode frequency. From Poynting theorem \cite{tobar2021abraham,TobarQEMD22} for a resonant system one can show that on resonance the imaginary term of Poynting theorem goes to zero (no reactance on resonance), and the power flow is real, which flows into the cavity through the dissipative channel given by
\begin{equation}
\begin{aligned}
&\oint\operatorname{Re}\left(\mathbf{S}_{1}\right)\cdot \hat{n}ds=\int\Big(-\frac{1}{4}(\mathbf{E}_1 \cdot\mathbf{J}_{e1}^*-\mathbf{E}_1^* \cdot\mathbf{J}_{e1})\\
&+\frac{j\omega_a\epsilon_0cg_{a\gamma\gamma}}{4}(\mathbf{E}_1 \cdot\tilde{a}^*\mathbf{B}_0^*
-\mathbf{E}_1^* \cdot\tilde{a}\mathbf{B}_0)\Big)~dV.
\end{aligned}
\label{PoyntTh}
\end{equation}
Since the phase of the photon leaving the cavity is not our observable, we may arbitrarily set the axion phase. Setting $a_0=\tilde{a}=\tilde{a}^*=\sqrt{2}\langle a_0\rangle$, Eqn.~(\ref{PoyntTh}) becomes
\begin{equation}
\begin{aligned}
&\oint\operatorname{Re}\left(\mathbf{S}_{1}\right)\cdot \hat{n}ds=\int\Big(-\frac{1}{4}(\mathbf{E}_1 \cdot\mathbf{J}_{e1}^*-\mathbf{E}_1^* \cdot\mathbf{J}_{e1})\\
&+\frac{j\omega_a\epsilon_0cg_{a\gamma\gamma}\sqrt{2}\langle a_0\rangle}{4}(\mathbf{E}_1 \cdot\mathbf{B}_0^*-\mathbf{E}_1^* \cdot\mathbf{B}_0)\Big)~dV.
\end{aligned}
\label{PoyntAC}
\end{equation}
Noting that the power generated by the axion is the last term in Eqn.~(\ref{PoyntAC}), and ignoring losses in the background field, so $\mathbf{B}_0$ is real, we find the signal power is given by
\begin{equation}
\begin{aligned}
P_{s1}=g_{a\gamma\gamma}\frac{\omega_a\epsilon_0c\langle a_0\rangle}{\sqrt{2}Q_1}&\int\operatorname{Re}(\mathbf{E}_1)\cdot\operatorname{Re}(\mathbf{B}_0) dV.
\end{aligned}
\label{PVDM2}
\end{equation}
In the steady state we equate $P_{s1}$ in (\ref{PVDM2}) to the dissipated power, $P_d=\frac{\omega_1U_1}{Q_1}$, then the stored energy is,
\begin{equation}
\begin{aligned}
U_1=g_{a\gamma\gamma}\frac{\omega_a\epsilon_0\langle a_0\rangle}{\sqrt{2}\omega_1}\int\operatorname{Re}(\mathbf{E}_1)\cdot\operatorname{Re}(c\mathbf{B}_0)~dV.
\end{aligned}
\label{StEngEB}
\end{equation}
This is essentially the same process as undertaken by Lasenby and Berlin et. al. \cite{berlin2020axion,Lasenby2020,Lasenby2020b,ABerlin2021}, which matches the signal power to the dissipated power, so all three independent calculations give essentially the same sensitivity. We take it a step further here by calculating the sensitivity including cavity parameters.

Since $U_1=\frac{\epsilon_0}{2}\int\mathbf{E}_1 \cdot\mathbf{E}_1^*dV$, we obtain,
\begin{equation}
\begin{aligned}
\sqrt{U_1}=g_{a\gamma\gamma}\frac{\omega_a\epsilon_0\langle a_0\rangle}{\sqrt{2}\omega_1}\frac{\int\operatorname{Re}(\mathbf{E}_1)\cdot\operatorname{Re}(c\mathbf{B}_0)~dV}{\sqrt{\frac{\epsilon_0}{2}\int\mathbf{E}_1 \cdot\mathbf{E}_1^*~dV}}.
\end{aligned}
\label{SqrtStEngEB}
\end{equation}
Then defining the unit vectors so $c\mathbf{B}_0=E_{00}\mathbf{b}_0$ and $\mathbf{E}_1=E_{01}\mathbf{e}_1$ Eqn.~(\ref{SqrtStEngEB}) becomes
\begin{equation}
\begin{aligned}
\sqrt{U_1}=g_{a\gamma\gamma}\frac{\omega_a\sqrt{\epsilon_0}\sqrt{V}E_{00}\langle a_0\rangle}{\omega_1}
&\frac{\frac{1}{V}\int\mathbf{e}_1\cdot\mathbf{b}_0~dV}{\sqrt{\frac{1}{V}\int\mathbf{e}_1 \cdot\mathbf{e}_1^*~dV}}\\
=g_{a\gamma\gamma}\langle a_0\rangle\frac{\omega_a}{\omega_1}\sqrt{\frac{2P_{c0}}{\omega_0}}\xi_{10},
\end{aligned}
\label{U1}
\end{equation}
where the overlap functions are defined by
\begin{equation}
\xi_{10}=\frac{1}{V}\int\mathbf{e}_1\cdot\mathbf{b}_0~dV~\text{and}~\xi_{01}=\frac{1}{V}\int\mathbf{e}_0\cdot\mathbf{b}_1~dV,
\label{Olap}
\end{equation}
where ($\mathbf{e}_0, \mathbf{b}_0$) and ($\mathbf{e}_1, \mathbf{b}_1$) are the mode electric and magnetic field real unit vectors \cite{Kakazu96,TobarQEMD22}, and the overlap functions are analagous to $\sqrt{ C_{01}}$, the square root of the form factor of a regular haloscope experiment, which is maximized when the cavity mode's electric field is parallel to the applied static magnetic field \cite{Bartram2021, Quiskamp2022}.

Here $E_{00}=\sqrt{\frac{2P_{c0}}{\omega_0\epsilon_0V}}$ 
where $P_{c0}$ is the circulating power of the background mode over the cavity volume $V$, which is related to the incident power, $P_{0inc}$, by
\begin{equation}
P_{c0}=\frac{4 \beta_{0} Q_{L0}}{(\beta_0+1)^2}P_{0inc},
\end{equation}
where $\beta_0$ is the coupling to the background mode and $Q_{L0}$ is the mode's loaded quality factor. Now, we can determine the square root of power in the coupling circuit of the readout mode to be
\begin{equation}
\sqrt{P_{1out}}=\frac{\sqrt{\omega_1Q_{L1}U_1}\sqrt{\beta_1}}{\sqrt{1+\beta_1}\sqrt{1+4Q^2_{L1}\big(\frac{\delta\omega_a}{\omega_1}\big)^2}},
\label{Pro}
\end{equation}
where we define $\delta \omega_a=\omega_a-|\omega_0-\omega_1|$, so when $\delta\omega_a=0$ then $\omega_a=\omega_0-\omega_1$ and the axion induced power is upconverted to the frequency, $\omega_1$. Thus, $\delta\omega_a$ defines the detuning of the induced power with respect to the readout mode frequency.
Combining (\ref{U1})-(\ref{Pro}), we obtain
\begin{equation}
\begin{aligned}
\sqrt{P_{1out}}=\mathcal{K}p_{a\gamma\gamma}~g_{a \gamma\gamma}\left\langle a_{0}\right\rangle,
\end{aligned}
\label{PowerSens}
\end{equation}
where the transduction coefficient  in units of square root power is defined by
\begin{equation}
\begin{aligned}
&\mathcal{K}p_{a\gamma\gamma}=\frac{\xi_{10}2\sqrt{2}\omega_a\sqrt{\beta_{0}Q_{L0}\beta_1Q_{L1}P_{0inc}}}{\sqrt{\omega_1\omega_0}\sqrt{1+\beta_1}(\beta_0+1)\sqrt{1+4Q^2_{L1}\big(\frac{\delta\omega_a}{\omega_1}\big)^2}}. \\
\end{aligned}
\label{PowerSens2a}
\end{equation}
Applying Poynting vector analysis in a similar way to QEMD (quantum electromagnetodynamics) \cite{TobarQEMD22}, it has been shown that the transduction coefficient to $g_{aBB}$ is,
\begin{equation}
\begin{aligned}
&\mathcal{K}p_{aBB}=-\frac{\xi_{01}2\sqrt{2}\omega_a\sqrt{\beta_{0}Q_{L0}\beta_1Q_{L1}P_{0inc}}}{\sqrt{\omega_1\omega_0}\sqrt{1+\beta_1}(\beta_0+1)\sqrt{1+4Q^2_{L1}\big(\frac{\delta\omega_a}{\omega_1}\big)^2}}.
\end{aligned}
\label{PowerSens2b}
\end{equation}
Thus, the power technique is sensitive to the effective monopole coupling term $g_{aBB}$ unlike the Sikivie-type detectors that utilises a DC $\vec{B}$ field.

To calculate the signal to noise ratio (SNR) for virialized axion dark matter from the galactic halo, we take into account that it presents as a narrowband noise source with a linewidth of one part in $10^6$. In SI units we may relate the axion amplitude to the background  dark matter density in the galactic halo, $\rho_a$, by $\left\langle a_{0}\right\rangle=\frac{\sqrt{\rho_{a} c^{3}}}{\omega_{a}}$. Limits on the axion couplings ($g_i$, $i = a\gamma\gamma$, $aBB$) can be found by calculating
\begin{equation}\label{SNR}
SNRp_{i}=g_i \frac{|\mathcal{K}p_{i}|}{\omega_{a}\sqrt{P_N}}\sqrt{\rho_{a} c^{3}}\left(\frac{t}{\Delta f_a}\right)^{\frac{1}{4}},
\end{equation}
where $P_N$ (W/Hz) is the noise power competing with the axion signal and $\Delta f_a$ is the axion bandwidth in Hz, where $\Delta f_a=\frac{f_a}{10^6}$ for virialized dark matter. This assumes the measurement time, $t$ is greater than the axion coherence time so that $t>\Delta f_a^{-1}$. For measurement times of $t<\frac{10^6}{ f_{a}}$ we substitute $\left(\frac{10^6t}{ f_{a}}\right)^{\frac{1}{4}} \rightarrow t^{\frac{1}{2}}$. The noise power in such experiments is dominated by thermal noise in the readout mode of effective temperature $T_1$ and the noise temperature of the first amplifier after the readout mode, $T_{amp}$, and is given by \cite{HARTNETT2011,parker2013b}
\begin{equation}
\begin{aligned}
P_N\sim\frac{4 \beta_1 }{(\beta_1 +1)^2 \left(1+4 Q_{L1}^2\big(\frac{\delta\omega_a}{\omega_1}\big)^2\right)} \frac{k_BT_{1}}{2}+\frac{k_BT_{amp}}{2}.
\end{aligned}
\label{PN}
\end{equation}
In the case $\beta_1\sim1$ and $\delta\omega_a\sim0$ then $P_N\sim\frac{k_B(T_1+T_{amp})}{2}$, and assuming $\beta_0\sim1$ the signal to noise ratios become,
\begin{equation}
\begin{aligned}
&SNR_{a\gamma\gamma}\sim g_{a\gamma\gamma}
|\xi_{10}|\sqrt{\frac{2Q_{L0}Q_{L1}P_{0inc}\rho_a c^3}{\omega_1\omega_0k_B(T_1+T_{amp})}}
\left(\frac{10^6t}{ f_{a}}\right)^{\frac{1}{4}}, \\
&SNR_{aBB}\sim g_{aBB}
|\xi_{01}|\sqrt{\frac{2Q_{L0}Q_{L1}P_{0inc}\rho_a c^3}{\omega_1\omega_0k_B(T_1+T_{amp})}}
\left(\frac{10^6t}{ f_{a}}\right)^{\frac{1}{4}}.
\end{aligned}
\label{SNRa}
\end{equation} 

\begin{figure}[t!]
\center
\includegraphics[width=0.95\columnwidth]{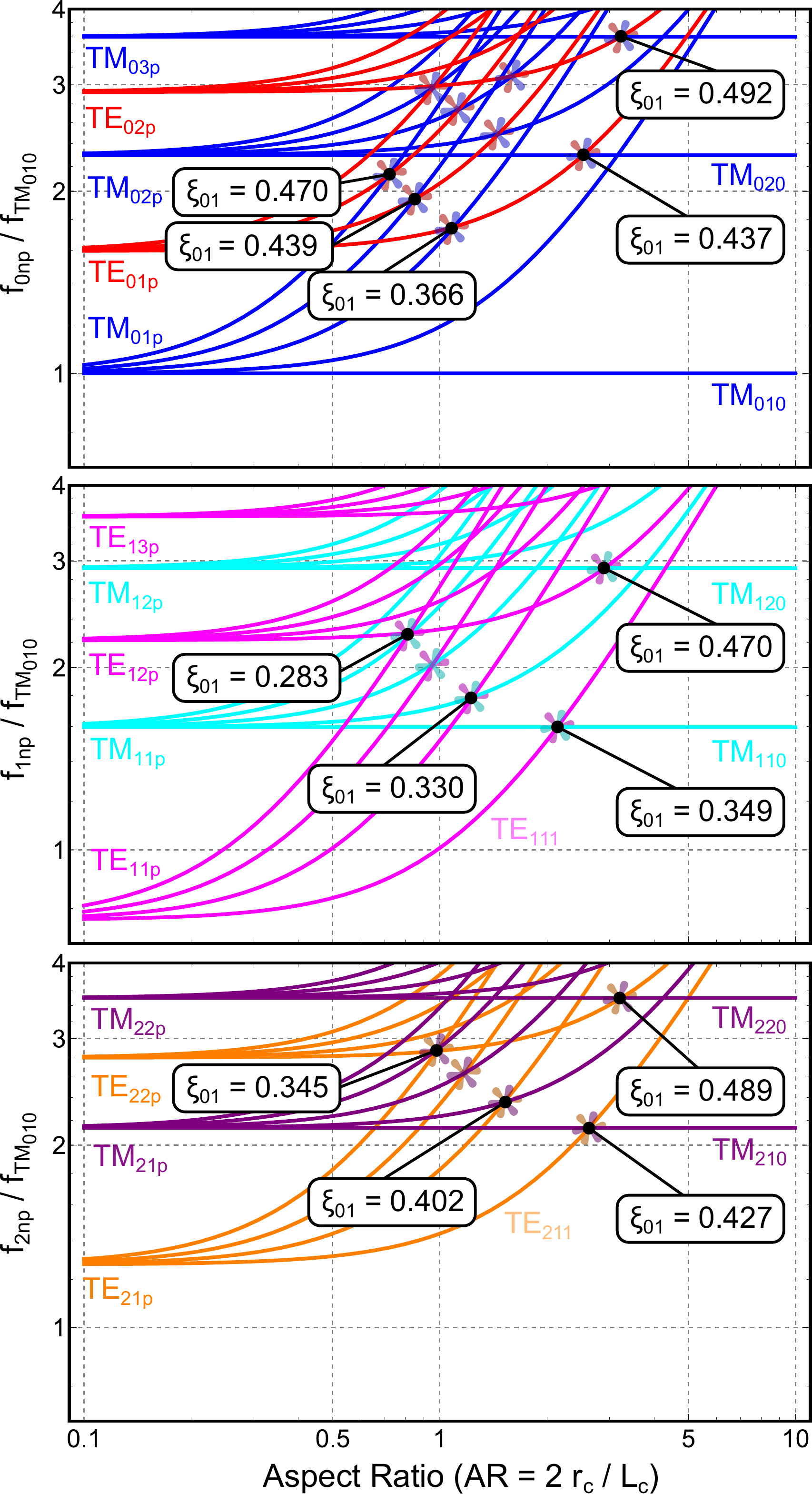}
\caption{Frequency mode chart for the $TM_{mnp}$ and $TE_{mnp}$ modes (normalized with respect to the $TM_{010}$ mode) versus aspect ratio, $AR=\frac{2r_c}{L_c}$. \textbf{Top:} $m=0$ and $p=0$ to $4$. \textbf{Center:} $m=1$ and $p=1$ to $4$. \textbf{Bottom:} $m=2$ and $p=1$ to $4$. Here $r_c$ and $L_c$ are the radius and length of the cavity respectively. Axion upconversion occurs when the two modes are close in frequency. The mode pairs with favourable overlap are highlighted with an asterisk, showing $\xi_{01}$ calculated when the mode frequencies coincide.}
\label{Olapm0}
\end{figure}

\subsection{Sensitive Mode-Pairs for Axion Upconversion Experiments in a Cylindrical Cavity}

\begin{figure}[t!]
\includegraphics[width=0.95\columnwidth]{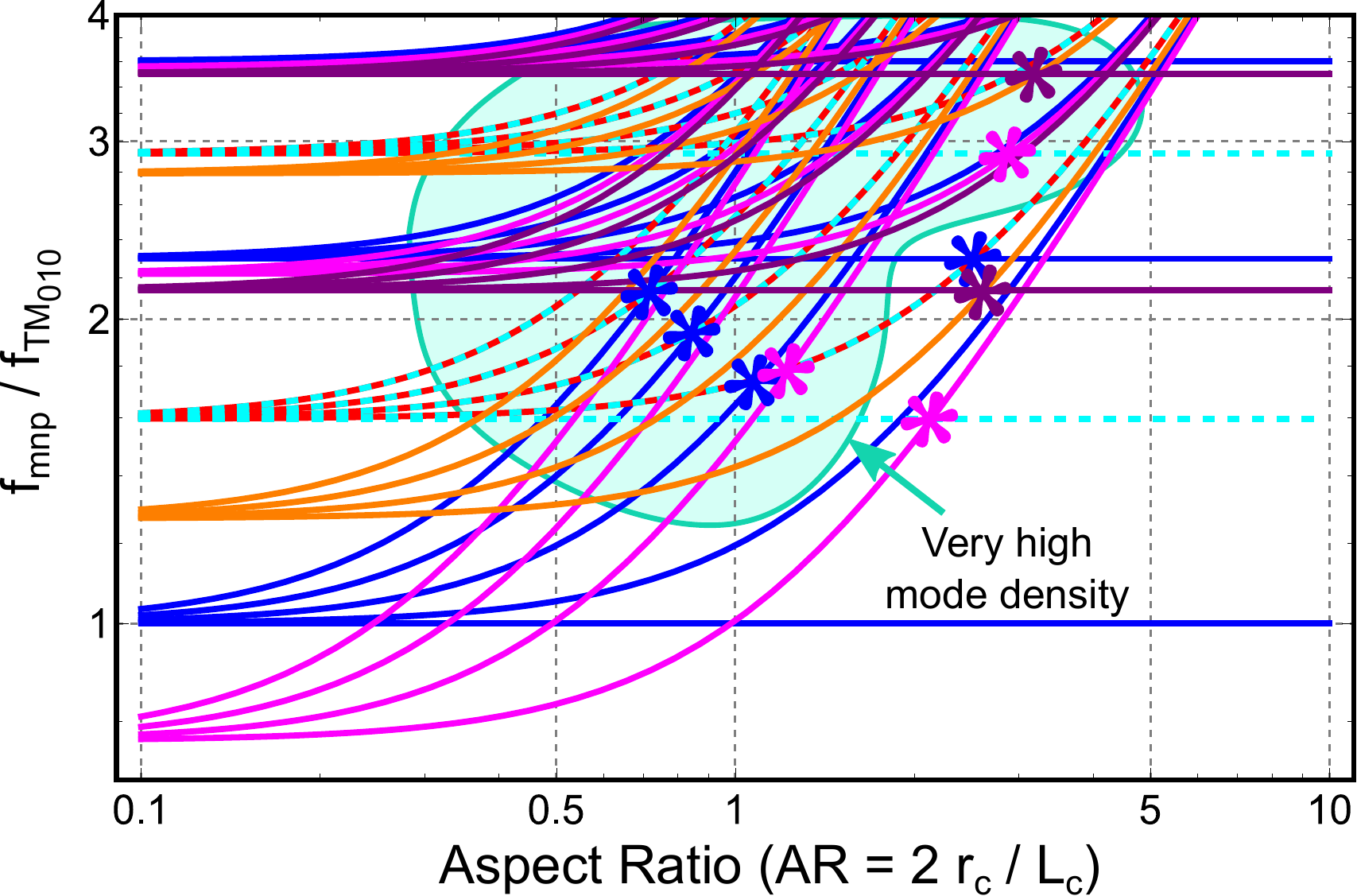}
\caption{Mode charts of Fig.~\ref{Olapm0} plotted together, with mode pairs of high overlap marked by asterisks. To keep the mode density low (and therefore avoid mode competition and mode crossings), the figure reveals that the an aspect ratio between 1 and 3 is optimal to achieve axion upconversion.}
\label{Olapm2}
\end{figure}

In general, modes that resonate in a cylindrical cavity may be classified as transverse magnetic (TM) or transverse electric (TE) with respect to cylindrical coordinates $(r,\phi,z)$. TM modes have no magnetic field in the $z$-direction ($B_z=0$), and TE modes have no electric field in the $z$-direction ($E_z=0$). A mode can be further defined by its standing wave field pattern in the cylindrical cavity, characterized by three numbers: $m$, the number of azimuthal variations, which must be greater than or equal to 0; $n$, the number of radial half-variations across the diameter, which must be greater than or equal to 1; and $p$, the number of axial half-variations, which must be greater than or equal to 0 for TM modes or greater than or equal to 1 for TE modes. The resultant mode is labelled either $TM_{mnp}$ or $TE_{mnp}$.

The choice of the two cavity modes for the experiment is dictated by their favourable electromagnetic overlap function in (\ref{Olap}) where a larger overlap function increases the sensitivity of the experiment to axion-photon couplings, $g_{a\gamma\gamma}$ and $g_{aBB}$. $\xi_{10}$ and $\xi_{01}$ may take a maximum value of 1 and Fig.~\ref{Olapm0} illustrates the non-zero overlap of certain pairs of TE and TM modes in an arbitrary right cylindrical cavity. For a high overlap coefficient, it appears that the first condition is that both modes should share the same azimuthal mode number $m$. On inspection, mode pairs with good overlap include, $TE_{011}-TM_{020}$ with $\xi_{01}=0.437$, $TE_{011}-TM_{012}$ with $\xi_{01}=0.366$, $TE_{012}-TM_{013}$ with $\xi_{01}=0.439$, $TE_{112}-TM_{111}$ with $\xi_{01}=0.330$ and $TE_{211}-TM_{210}$ with $\xi_{01}=0.427$. These modes are plotted together in Fig.~\ref{Olapm2}, and we notice the best mode pairs differ in value of axial mode number $p$ by 1. A further complication occurs if the azimuthal mode number, $m$, is non-zero. In this case, there may be a phase shift between the azimuthal nodes of the two modes and thus the sensitivity could be severely reduced. So, for practical reasons it is best to consider mode pairs with $m=0$, eliminating the need to confirm the spatial dependence of the degenerate modes. In the end we chose to use the $TE_{011}-TM_{020}$ mode pair for our experiment.

Unit vectors for the electric and magnetic field components of TE and TM modes have analytical solutions for the cylindrical cavity. As such, $\xi$ can be found analytically via Eqn.~(\ref{Olap}). Appendix~\ref{appendix:B} derives the following result for arbitrary $TM_{0,n,0}$ and $TE_{0,n,p}$ modes:

\begin{equation}
\begin{split}
&\xi_{01}=\frac{1}{V}\int\mathbf{e}_{n_0p_0}\cdot\mathbf{b}_{n_1}dV=\frac{4\sqrt{2}\chi_{0n_0}^\prime}{p_0\pi(\chi_{0n_0}^{\prime 2}-\chi_{0n_1}^2)},\\
&\xi_{10}=\frac{1}{V}\int\mathbf{e}_{n_1}\cdot\mathbf{b}_{n_0p_0}dV=\frac{4\sqrt{2}\chi_{0n_0}^\prime}{p_0\pi(\chi_{0n_0}^{\prime 2}-\chi_{0n_1}^2)}\frac{\omega_1}{\omega_0},
\end{split}
\label{Olap01example}
\end{equation}

where $\chi_{a,b}$ represents the $b^{th}$ root of the $a^{th}$ Bessel function, $\chi_{a,b}'$ represents the $b^{th}$ root of the derivative of the $a^{th}$ Bessel function, $\omega_0$ is the pump mode frequency and $\omega_1$ is the readout mode frequency.

\subsubsection{Sensitivity of the $TE_{011}-TM_{020}$ mode pair}

In this work we implement the $TE_{011}-TM_{020}$ mode pair where the frequency of the $TM_{020}$ pump mode is first order insensitive to the tuning mechanism, and is stationary at 8.99 GHz, while the readout mode is tuned between 8.70 to 8.72 GHz to search for axions from 270 MHz to 290 MHz. For these tuning ranges $\xi_{01}=$-0.437 and $\xi_{10}$ varies from -0.451 to -0.449.

\section{Comparison of the Frequency and Power Techniques}

As previously stated, two methods, the ``frequency technique'' and the ``power technique'' have been proposed and tested in this work. 

We have recently shown that when one compares dissimilar axion haloscopes the use of spectral density of photon-axion theta angle noise is a practical comparison parameter to use \cite{sym14102165}, and is given by (here subscript 0 represents the background mode and 1 the readout mode),
\begin{equation}
\begin{aligned}
&\sqrt{S_{\theta}}=\frac{\sqrt{\omega_0\omega_1}\sqrt{k_B}(1+\beta_0)}{2\xi_{10}\omega_a\sqrt{2}\sqrt{\beta_1Q_{L1}\beta_0Q_{L0}P_{0_{inc}}}}\times \\
&\sqrt{\frac{2T_{1}\beta_1}{(\beta_1+1)}+\frac{T_{RS}(\beta_1+1)}{2}\left(1+4 Q_{L1}^2\big(\frac{\delta\omega_a}{\omega_1}\big)^2\right)},
\end{aligned}
\label{SpecSens}
\end{equation}
for both the frequency and power upconversion techniques. Here, $P_{0_{inc}}$ is the power incident on the cavity pump mode, $\beta_0$ and $\beta_1$ are the cavity probe couplings, $Q_{L0}$ and $Q_{L1}$ are the loaded quality factors, $k_B$ is the Boltzmann constant and $\delta \omega_a=\omega_a-|\omega_1-\omega_0|$ is the angular frequency offset of axion signal from readout mode resonance. $T_1$ and $T_{RS}$ are the haloscope cavity temperature, which determines the Nyquist noise generated in the cavity, and the added readout system noise temperature, respectively. Eqn.~(\ref{SpecSens}) leads to the signal to noise ratio for a signal of virialized galactic halo dark matter axions to be given by \cite{sym14102165},
\begin{equation}
\sqrt{SNR}= g_{a\gamma\gamma}  \frac{\sqrt{\rho_a c^{3}}~t^{\frac{1}{4}}\left(\frac{10^6}{f_a}\right)^{\frac{1}{4}}}{\omega_a\sqrt{S_{\theta}}},
\label{SNRgen}
\end{equation}
where $\rho_a$ is the galactic halo dark matter density, and $t$ the measurement time assuming $t>\tau_a=\frac{10^6}{f_a}$, where $\tau_a$ is the axion coherence time.

For the power technique, $T_{RS}$ will be dominated by the noise temperature of the first amplifier in the readout chain, while for the frequency technique it will be dominated by the noise temperature of the phase noise suppression system. For a well-designed system this should be equal to the noise temperature of a low noise amplifier in the phase detection circuitry. Therefore, the sensitivity of both methods should be essentially the same. Experimentally, the frequency method poses greater technological challenges due to the need to reject noise in the dual-pumped resonator. In this work where we search for axion masses with equivalent frequencies between 270 and 290 MHz, we found the power method a more robust experiment to achieve the Nyquist limit, and so this method is reported below. The reader may nevertheless find the experimental details of the dual-pumped ``frequency method'' interesting to read about in Appendix \ref{appendix:A}.

\begin{figure}[t]
\includegraphics[width=1.0\columnwidth]{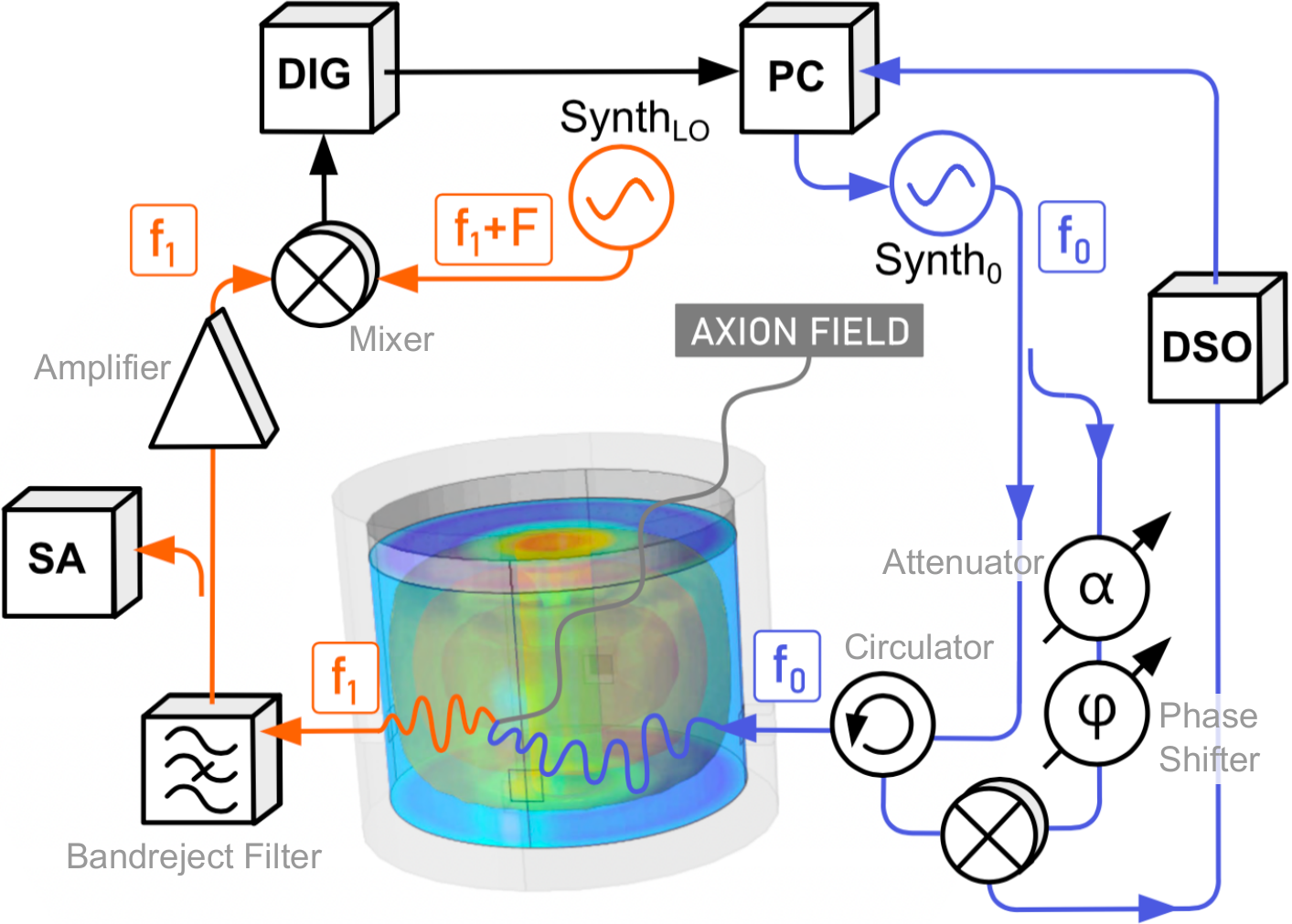}
\centering
\caption{Schematic of the power experiment. A synthesizer provides 15 dBm of incident power to the pump mode ($\text{TM}_{020}$) which interacts with the axion field ($f_a \approx | f_1 - f_0 |$) to deposit power within the bandwidth of the readout mode at $f_1$ ($\text{TE}_{011}$). A bandreject filter suppresses parasitic feedthrough of $f_0$ such that $f_1$ is more efficiently amplified before being mixed down to $F$ (arbitrary, $\sim78$ MHz) and digitized (DIG). The pump tone ($f_0$) is kept on resonance with the $\text{TM}_{020}$ mode by periodically sweeping the tone and viewing the response via a phase bridge connected to a digital storage oscilloscope (DSO).}
\label{PowerMethod}
\end{figure}

\section{Room Temperature Power Experiment}
\begin{figure}[b]
\includegraphics[width=1\columnwidth]{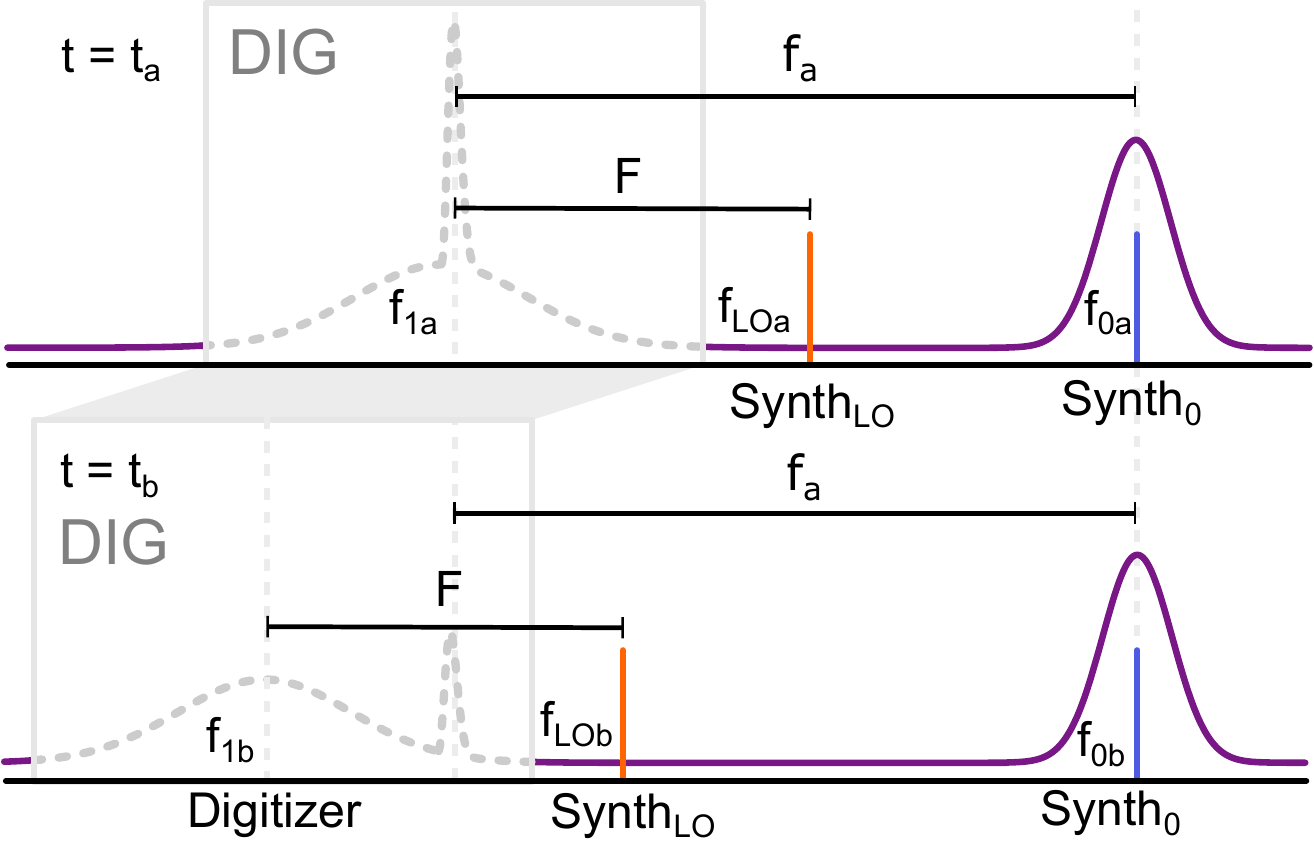}
\centering
\caption{Data is collected at the digitizer (DIG) at a given offset ($f_a \approx | f_1 - f_0 |$) for approximately 8 hours before $f_1$ is tuned via cavity height in order to scan an adjacent region of axion space. IF (intermediate frequency) interference which is stationary with respect to $f_{LO}$ (secondary synthesizer frequency i.e. local oscillator) is not stationary with respect to axion space. Therefore, such interference is not coherently combined with the grand spectrum and can be easily vetoed from the axion candidate list.}
\label{tuning}
\end{figure}

Fig.~\ref{PowerMethod} depicts the experimental setup for the power upconversion experiment. The cavity is a closed height-tunable silver-plated copper cylinder of radius 29.3 mm, supporting a pump mode ($\text{TM}_{\text{020}}$, $f_0$) at 8.99 GHz and a readout mode ($\text{TE}_{\text{011}}$, $f_1$) mode which can tune from 6.66 GHz to 12.13 GHz by varying the cavity height between 6.4 cm and 1.4 cm via a micrometer attached to the cavity lid.

A probe oriented to couple to the pump mode is provided with 31 mW (15 dBm) of power at its resonant frequency (from synthesizer 0), and a probe oriented to couple to the readout mode collects data to examine for evidence of axion upconversion. This experiment is very similar to a regular (DC) haloscope, such as ADMX \cite{Bartram2021} and ORGAN \cite{Quiskamp2022} with the exception that the pump photon is sourced from a second mode within the cavity, instead of from a DC magnet surrounding the cavity.

The axion interacts as a narrowband noise source with a frequency of $f_a$. If our experiment is tuned such that the axion frequency satisfies $f_a = f_1 - f_0$ then we expect for this axion source to be upconverted via the strong pump mode ($f_0$) into photons at the readout frequency ($f_1$). In our analysis, we assume that dark matter follows the standard virialized halo model, and as such that the relative axion velocity distribution is Maxwell-Boltzmannian, with an rms velocity of  $v_c = 226 \text{ km s}^{-1}$ \cite{Lentz2017}. Thus, data analysis consists of searching for a narrow peak above the thermal noise of $f_1$ resembling the expected axion lineshape.

The axion-induced amplitude noise (W/Hz) would appear in the power spectral density of amplitude noise of the readout mode in the form

\begin{equation}
    \begin{split}
  & S_{A\alpha_{1}}(f_a) = g_{a\gamma\gamma}^2k_{a}^2S_A(f_a), \\
  &k_{a}= \frac{f_a}{\sqrt{f_1 f_0}} \frac{2 \sqrt{2} \sqrt{\beta_1 \beta_0}\sqrt{Q_{L1} Q_{L0}  P_{0_{inc}}}\xi_{10}}  {\sqrt{1+\beta_1}(1+\beta_0)\sqrt{1+4 Q_{L1}^2\left(\frac{f_a-|f_0-f_1|}{f_1}\right)^2}} 
  \end{split}
    \label{AxionPSD}
\end{equation}

where $\xi_{10}$ is the geometric overlap factor, describing the efficiency of photon upconversion between the two electromagnetic modes. Here $k_{a}$ is the conversion ratio from axion theta angle, $\theta=g_{a\gamma\gamma} S_A(f_a)$, to amplitude noise. As previously noted, the axion (or axion-like particle) field may be considered as a spectral density of narrowband noise, centred at a frequency equivalent to the axion mass and broadened due to cold dark matter virialization to give a linewidth of approximately ${10^{-6}{f_{a}}}$, and is denoted as ${S_{A}(f)}$ (kg/s/Hz) \cite{Lentz2017}. This signal must compete against the combined Nyquist noise of the readout mode and the thermal noise of the readout amplifier (with units of W/Hz) which is given by
\begin{equation}
\begin{split}
S_{N\alpha_{1}}(f_a) = & \frac{4 \beta_1 }{(\beta_1 +1)^2 \left(1+4 Q_{L1}^2\big(\frac{f_a-|f_0-f_1|}{f_1}\big)^2\right)} \frac{k_BT_{1}}{2}\\
&+\frac{k_BT_{amp}}{2}
\end{split}
\label{thermal}
\end{equation}

\begin{figure*}
\includegraphics[width=2\columnwidth]{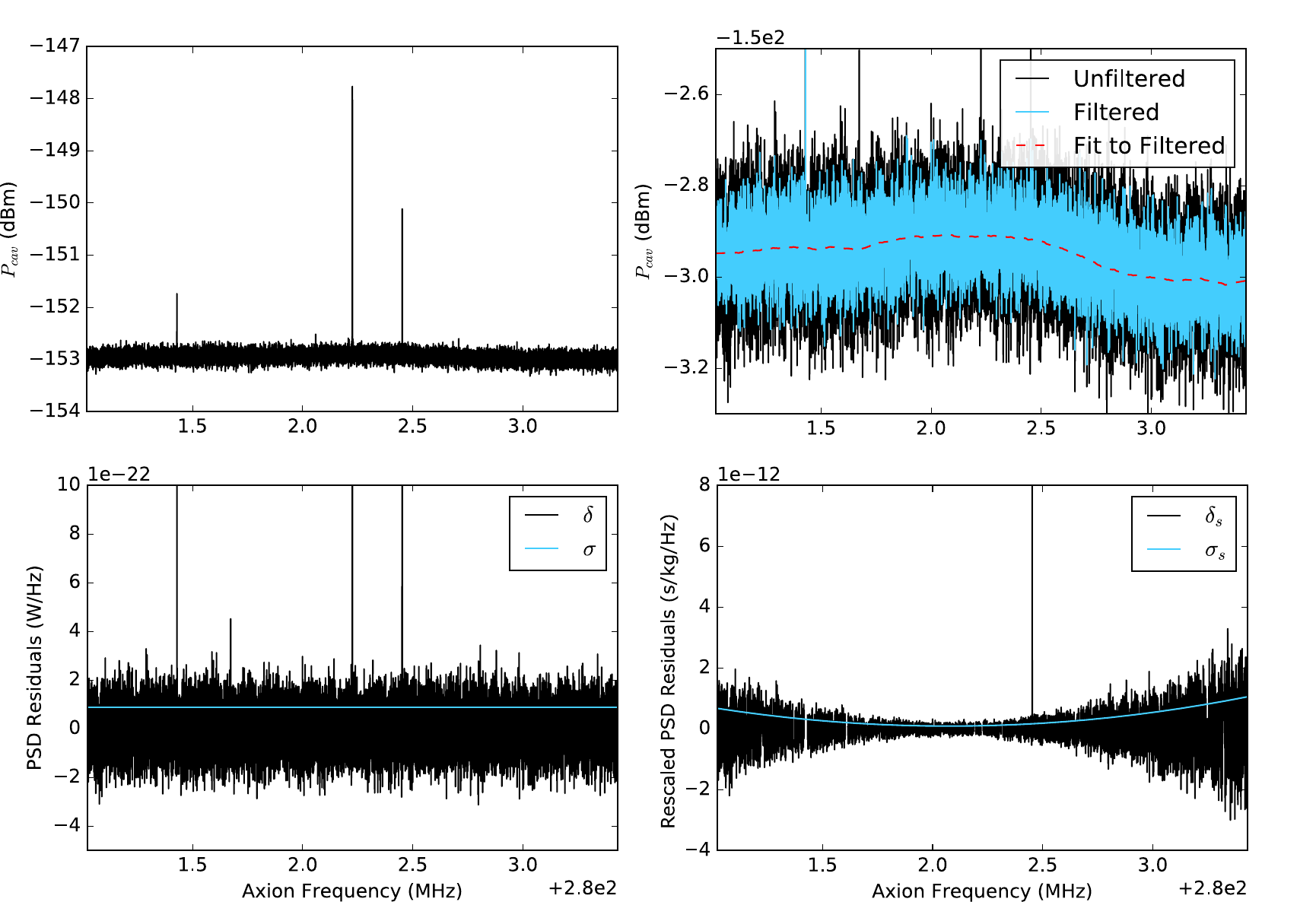}
\centering
\caption{\textbf{Top left:} Power coupled into the receiver chain from the thermal noise of the TE mode. \textbf{Top right:} Discriminating a baseline via peak rejection and Savitzky-Golay filtering. Peak rejection before filtering minimizes the potential of an axion signal skewing the baseline in a manner detrimental to SNR. \textbf{Bottom left:} Background subtracted power spectral density, leaving us with a forest of residuals within which to discriminate an axion signal. \textbf{Bottom right:} Residuals scaled by the conversion coefficient $k_a$, such that an axion of coupling strength $g_{a\gamma\gamma}$ would appear at the same amplitude in this spectrum regardless of frequency. One can clearly see that SNR is maximized on mode resonance.}
\label{processing}
\end{figure*}

\begin{figure}[t]
\includegraphics[width=1\columnwidth]{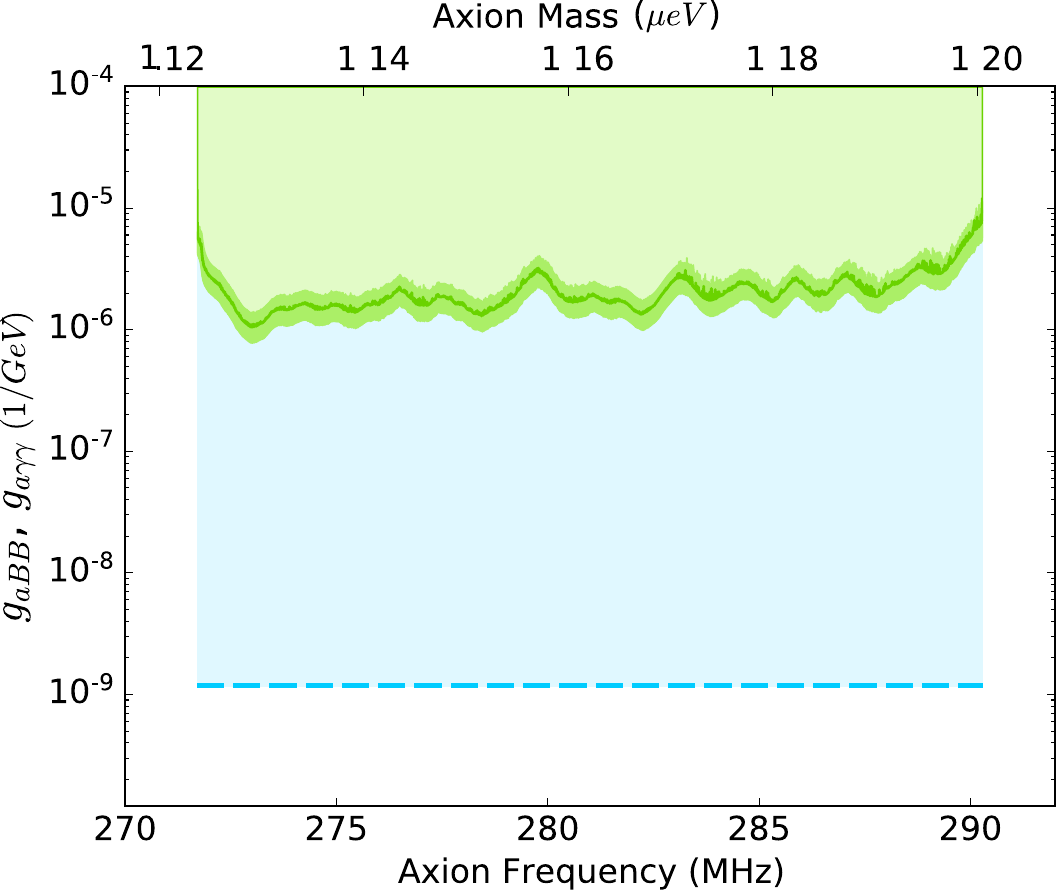}
\centering
\caption{In green, the 95\% confidence axion exclusion zone for both $g_{a\gamma\gamma}$ and $g_{aBB}$ for the measured mass range between 1.12 $-$ 1.20 $\mu eV$ (271.7 MHz - 290.3 MHz) for a measurement period of 30 days, which is a three order of magnitude improvement over our previous result \cite{Cat21}. The bright green region represents the uncertainty on excluded $g_{a\gamma\gamma}$ which is detailed in Appendix~\ref{appendix:C}. The blue dashed line represents the approximate sensitivity achievable with a niobium resonator of loaded quality factors around $10^7$ and cooled to a temperature of 4~K, measuring for a period of 30 days, and using a cryogenic amplifier of noise temperature 4~K. Construction for this setup is underway.
 }
\label{exclusionlimits}
\end{figure}
where $T_1$ is the temperature of the cavity and $T_{amp}$ is the effective temperature of the readout amplifier.

Synthesizer 0 provides the cavity with power at a frequency of $f_0$, which is on resonance with the TM mode. Whilst this mode is height-insensitive, and therefore unresponsive to the tuning mechanism, its frequency nevertheless wanders due to temperature fluctuations and in response to mode crossings. Therefore, synthesizer 0 is programmed to follow this wandering such that $f_0$ remains approximately on resonance. Analogue active feedback to the modulation port of the synthesizer was initially used for frequency locking, however, we found that the jitter of the pump mode during a data acquisition period introduced an unnacceptable and unnecessary level of uncertainty with respect to axion frequency. Therefore, real-time active feedback was abandoned and replaced with pump-tuning steps between measurements, occuring every minute. Of course, $f_0$ will consequently not be precisely on resonance with the pump mode at all times. Nevertheless, the uncertainty in pump power impacts SNR far less than uncertainty in axion frequency. During a pump-tuning step, the difference between the cavity resonance and the current synthesizer frequency is determined via frequency sweeping $f_0$ and observing the amplitude of reflection via an oscilloscope at the output of a phase bridge set to amplitude sensitivity. Thus $f_0$ is adjusted to follow the $TM_{020}$ resonance.

The readout channel includes a bandreject filter tuned to absorb power at $f_0$ which feeds through the readout probe, in order not to saturate the low noise amplifier which amplifies $f_1$ to a measureable level. The amplified noise is then mixed down with a second synthesizer which outputs a frequency of $f_{LO} =  f_1 + F$ where the frequency of $f_1$ is approximately stationary and manually tuned every eight hours. Therefore, digitizing and viewing a window around $F$ (a point in the IF spectrum, arbitrarily chosen to be 78 MHz) allows us to view the readout resonance.

Spectra of bin width 119 Hz and 3 MHz span (26,214 points), centralized on the readout mode, are averaged for 17 seconds before being saved to file. Each file also includes a record of $f_0$ and $f_{LO}$ at the time of measurement, such that $f_1$ (and therefore $f_a$) can be inferred. These power spectra are then aligned to the grand axion frequency spectrum, and scaled according to putative axion sensitivity ($k_a$) before being vertically combined with the grand spectrum via a maximum likelihood weighted average based on the inverse variance for a given bin. This process is detailed in the next section.

As $f_{LO}$ and $f_0$ are discretely tuned at each height tuning step and pump-tuning step respectively, each bin in the grand spectrum is a linear combination of contributions associated with different bins in the digitized Fourier spectrum (see Fig.~\ref{tuning}). Therefore, IF noise which is stationary with respect to $f_{LO}$ is not coherently added to the grand spectrum due to the fact that it is not stationary with respect to $f_a$. Conversely, axion-induced noise is coherently added. Tuning by height ($f_1$) or by LO frequency are thereby methods by which to confirm or veto axion candidates.

\subsubsection{Data Analysis}
The data anlysis procedure, which has been briefly outlined in the previous section, is based upon the HAYSTAC analysis procedure \cite{brubaker2017} and is visualized in Fig.~\ref{processing}. Firstly, we collect data from the output of the readout mixer with a National Instruments Digitizer (NI-5761R) with software written for the field-programmable gate array (NI-7935R) on LabVIEW by Paul Altin \cite{Quiskamp2022}. This observed power ($P_{obv}$), which has been filtered via the bandreject filter and amplified by the low noise amplifier is transformed to power at the output of the readout probe ($P_{cav}$) by subtracting the gain in the readout circuit. This data is translated to its appropriate frequency position in axion space via auxiliary data ($f_0$ and $f_{LO}$). Synthesizer 0 is always set to frequency values such that the edges of digitizer bins match exactly with the edges of the grand spectrum bins of the dataframe that is continuously updated with new data.
\begin{table}
\centering
\caption{Coupling and Loaded Quality Factors}
\label{tab:betaandQ}
\begin{tabular*}{\linewidth}{l@{\extracolsep{\fill}}lll}
\toprule
        	  & \text{Min}  & \text{Max}   \\ \midrule\midrule
$\beta_0$  &   0.762      & 0.775\\ 
$\beta_1$  &   0.867      & 0.891       \\ 
$Q_{L_0}$  &   6708      & 6720  \\ 
$Q_{L_1}$  &   10400       & 10418      \\  \bottomrule
\end{tabular*}
\end{table}

The baseline of $P_{cav}$ is found via Savitzky-Golay filtering a peak-eliminated spectrum which is then subtracted from $P_{cav}$ and divided by the resolution bandwidth to produce a power spectral density of residual noise ($\delta$, W/Hz). The peak of the baseline is recorded to infer $f_1$, the thermal TE mode resonance. The quality factors and couplings of the pump and readout modes at the tuning height are then extracted from the relevant callibration curves so that the conversion coefficient $k_a$ (Eqn.~(\ref{AxionPSD})) can be determined for each bin (see Table~\ref{tab:betaandQ} for an overview of quality factors and coupling coefficients).

The spectral density of raw residuals is divided by the spectrum of $k_a^2$ to produce the spectrum of scaled residuals ($\delta_s$). Specifically, raw cavity power spectral density in bin $i$ in the $j$-th measurement, corresponding to an axion frequency of $f_{a_{i,j}}$, is related to the relevant bin in the spectrum of scaled residuals via

\begin{equation}
\begin{split}
\delta_{s_{i,j}} =& \delta_{i,j} \Biggl( \frac{f_{a_{i,j}}}{\sqrt{f_{1_j} f_{0_j}}}\\
& \frac{2 \sqrt{2} \sqrt{\beta_{1_j} \beta_{0_j}}\sqrt{Q_{{L1}_j} Q_{{L0}_j} \xi P_{{0_{inc}}_j}}}  {\sqrt{1+\beta_{1_j}}(1+\beta_{0_j})\sqrt{1+4 Q_{{L1}_j}^2\left(\frac{f_{a_{i,j}}-|f_{0_j}-f_{1_j}|}{f_{1_j}}\right)^2}} \Biggr)^{-2}.
\label{deltas}
\end{split}
\end{equation}

Refering to Eqn.~(\ref{AxionPSD}), it is clear that an axion with a coupling constant of $g_{a\gamma\gamma}$ would appear in the spectrum of $\delta_{s}$ with an amplitude of $g_{a\gamma\gamma}^2 S_A(f_a)$ (kg/s/Hz) where $S_A(f_a)$ takes into account the Maxwell-Boltzmannian lineshape. The spectrum of standard deviations of the scaled residuals, $\sigma_s$ is similarly related to the standard deviation of the power spectral density of raw residuals, $\sigma$. Neglecting binning effects, an axion of a given $g_{a\gamma\gamma}$  would thus appear at the same amplitude in any bin of $\delta_s$ regardless of its position relative to the resonance peak.

As scaled residual spectra are collected, they are combined with the grand search spectrum ($\delta_g$) via a weighted average where each bin is weighted by its scaled inverse variance. Therefore, the amplitude of a bin $i$ in the grand search spectrum is equal to the sum of all weighted scaled contributions divided by the sum of all weights:

\begin{equation}
\delta_{g_{i}} = \frac{\sum_{j} \delta_{s_{i,j}} \left(\sigma_{s_{i,j}}\right)^{-2} }{\sum_{j} \left(\sigma_{s_{i,j}}\right)^{-2} }.
\label{dletag}
\end{equation}

Note that the weighted average preserves the units of $\delta_s$ (kg/s/Hz). The grand search spectrum standard deviation ($\sigma_g$) of a given bin $i$ is simply the inverted square root of the sum of all weights.

\begin{equation}
\left(\sigma_{g_{i}}\right)^{-2}  = \sum_{j} \left(\sigma_{s_{i,j}}\right)^{-2} 
\label{sigmag}
\end{equation}

$\sigma_{g_i}$ is an important parameter since it characterizes on a bin-by-bin basis the noise against which an axion signal is competing. Since the residual noise has a Gaussian distribution we set a $4 \sigma$ candidate threshold to flag axion candidates.

Sensitivity to axion peaks is enhanced by optimally filtering the search spectrum for the expected axion lineshape. Essentially, the spectrum is convolved with the axion lineshape such that axion-like signals are amplified in the data. Via Monte Carlo simulation, we found that an axion in the optimally filtered data with a $g_{a\gamma\gamma}$ corresponding to $9.5\sigma_g$ was discriminable above the candidate threshold with 95 percent confidence.

This detection efficiency could be somewhat improved with finer frequency resolution in the Fourier spectrum, however hardware limitations bottlenecked our bin width at 119 Hz. In our data, on average, the axion lineshape covers only 2.35 bins (with an axion bandwidth of approximately 280 Hz at $f_a =$ 280 MHz). Our matched filter therefore imparts a limited improvement to our detection efficiency compared to experiments with a higher number of bins per axion linewidth. Therefore, in Fig.~\ref{exclusionlimits}, axion exclusion limits are placed on $g_{a\gamma\gamma}$ (and $g_{aBB}$) consistent with the final amplitude of $9.5\sigma_g$ at the end of data processing.

\section{Prospects of detecting Low-Mass Axions using Upconversion}

Eqns.~(\ref{SpecSens}), (\ref{AxionPSD}) and (\ref{thermal}) posit that stronger axion exclusion limits could be set by the reported technique by selecting a resonator supporting suitable modes with higher quality factors at lower carrier frequencies (and hence larger volume), and also cooled to lower temperatures. Such suitable resonators have been reported in \cite{Martinello2018,Romanenko20,Posen20,Berlin2022} for 1 GHz cavities cooled to millikelvin temperatures. The quality factors of the modes in the cavity for our reported experiment were in the region of 6000 to 10,000, as expected from a silver-plated copper cavity. Superconducting niobium cavities, with a transition temperature of 9 K, offer a promising choice to further this technique, requiring however a greater level of engineering than the room temperature prototypes reported in this work, as detailed in recent proposals showing the potential of this idea~\cite{berlin2020axion,Lasenby2020b,ABerlin2021,Lasenby2020}. Such an experiment could put significant limits on the axion coupling term  $g_{aBB}$, to which the DC haloscope is insensitive.

\vspace{5mm}

\section*{Acknowledgments}

This work was funded by the Australian Research Council Centre of Excellence for Engineered Quantum Systems, CE170100009 and  Centre of Excellence for Dark Matter Particle Physics, CE200100008.

\appendix
\section{Prototype Frequency Experiment}
\label{appendix:A}

The experimental design of an improved frequency metrology experiment, a prototype of which was first reported in \cite{Cat21} is detailed in this section. The cavity, a cylindrical resonator made of silver-plated super invar, had a stationary $TM_{020}$ mode at $11.195$ GHz, and a $TE_{011}$ mode tuneable over a large range, with both loaded Q-factors of order $2500$. The Q-factor was degraded compared to our copper cavities, which may be due to an insufficient thickness of silver plating on  the invar cavity which has poor conductivity, a factor of 50 worse than copper. The decision to use this cavity was based upon the superior thermal expansion coefficient of super invar, however the suspension mechanism of the cavity lid, consisting of a commerical micrometer, highly dominated the long-term dimensional instability of the cavity, and therefore produced similar long-term thermal frequency oscillation in both the copper and invar cases.

Fig.~\ref{FreqMethod} depicts the frequency method experimental setup, which is pumped by a low-phase noise synthesizer. A signal reflected from the readout resonance was mixed down to baseband and amplified before its Fourier spectrum was collected and interrogated for axion-induced frequency noise. The challenge with this setup was maintaining perfect frequency lock with the readout mode and maximally suppressing its frequency noise, all the while tuning the cavity height at every tuning step.

Hybrid couplers paired with phase shifters and attenuators were used to achieve carrier suppression (via destructive interference) of both the parasitic feedthrough of the pump mode and the carrier of the readout mode (preserving sidebands) before amplification in the readout arm. It was necessary to suppress any pump mode feedthrough before amplification as not to saturate the readout amplifier. Suppression of the readout carrier serves a similar purpose, enabling a high frequency noise to voltage conversion efficiency without overdriving amplifier and mixer ports. This carrier-supressed readout mode was mixed down to baseband and collected via Fast Fourier Transform (FFT) at the vector signal analyzer. A portion was also filtered and sent to the modulation port of the synthesizer which pumps the readout mode. Thus, fluctuations of the target resonator mode ($f_1$) were actively followed by the synthesizer. Such frequency locking significantly improves the background noise level in the Fourier spectra which we collected for axion interrogation.

\begin{figure}[t!]
\includegraphics[width=1\columnwidth]{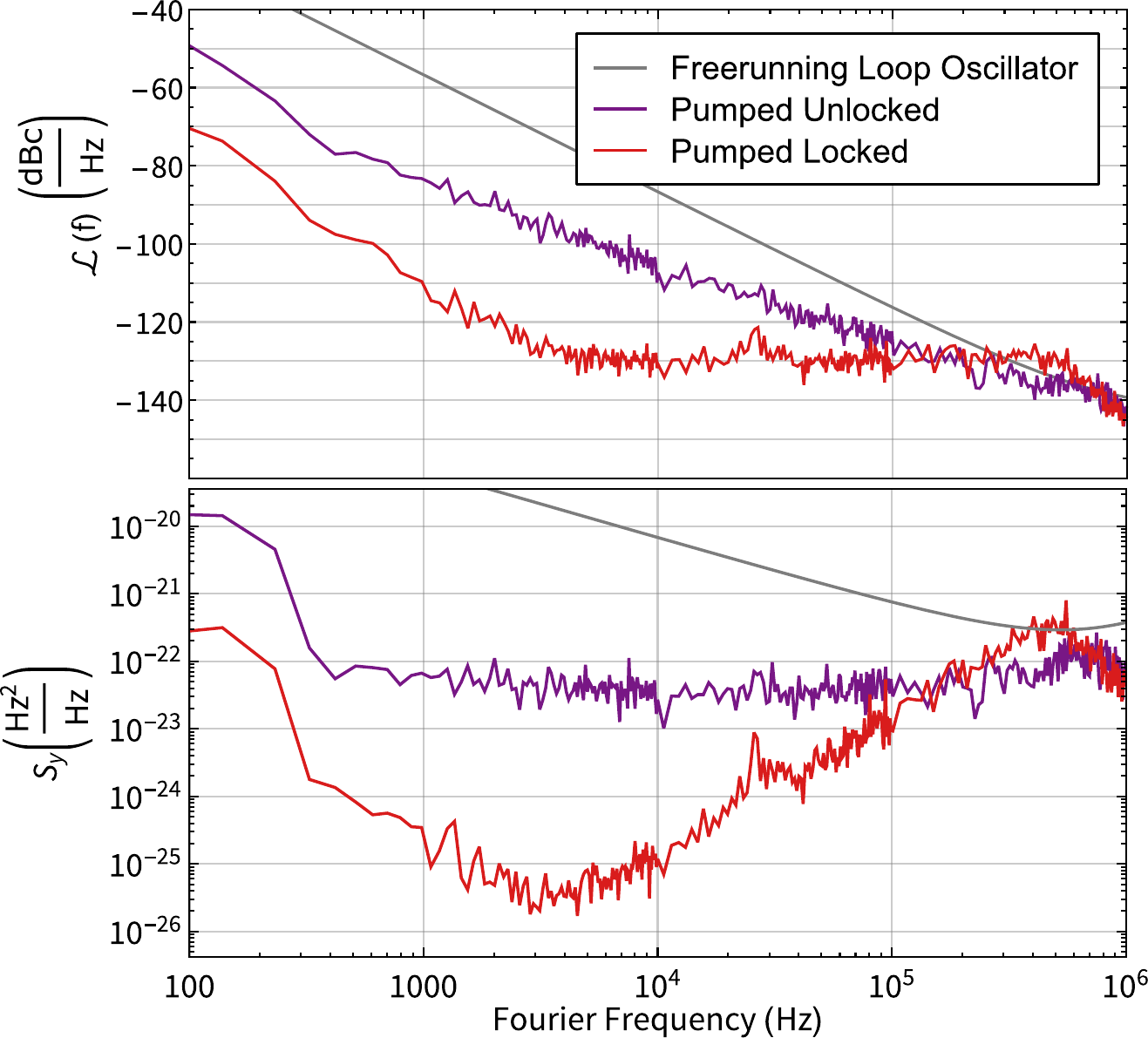}
\centering
\caption{Comparison between the freerunning loop oscillator scheme \cite{Cat21} and the pumped scheme of Fig.~\ref{FreqMethod}, with frequency locking enabled (active feedback to the synthesizer modulation port) and disabled. \textbf{Top:} Background noise represented as single sideband phase noise. \textbf{Bottom:} Background noise represented as fractional frequency noise.}
\label{Comparison}
\end{figure}

\begin{figure}[b!]
\includegraphics[width=1.0\columnwidth]{PumpedExperimentSchematic_Version6.pdf}
\caption{The frequency noise readout system. A pump frequency (blue, $f_0$) matching the $\text{TE}_{011}$ mode and a readout frequency (orange, $f_1$) matching $\text{TM}_{020}$ mode are provided via a synthesizer and a frequency shifter referenced to an arbitrary waveform generator (AWG). The reflection of the readout frequency from the cavity carries frequency noise information which is eventually discriminated and collected at baseband via Fast Fourier Transform (FFT). A digital storage oscilloscope (DSO) allows tuning of the carrier suppression system for the readout mode, consisting of manual and voltage-controlled phase shifters and attenuators (VCP and VCA) and two hybrid couplers which supress the carriers of the readout frequency and pump frequency respectively, improving the efficiency of frequency to voltage conversion at the readout mixer. A fraction of the frequency fluctuation information is filtered and returned to the synthesizer, such that the sythesizer remains locked on resonance with the TM mode. The beat note between $f_0$ and $f_1$ is recorded to align collected spectra in axion space in post-processing ($f_a = | f_1 - f_1 |$).}
\label{FreqMethod}
\end{figure}

Fig.~\ref{Comparison} illustrates the improvement in background noise over our previously published loop oscilllator experiment \cite{Cat21}, where $\sqrt{S_{y_1}}$ is the fractional frequency noise (Hz/$\sqrt{\text{Hz}}$) and may be theoretically predicted by

\begin{equation}
\sqrt{S_{y_1}}=\frac{\sqrt{k_{B} T_{R S}}}{\sqrt{2} Q_{L1} \sqrt{P_{1inc}}} \frac{(1+\beta_1)}{2 \beta_1}\sqrt{1+4Q_{L1}^2\left(\frac{\delta\omega_a}{\omega_{1}}\right)^{2}}.
\label{sigma2}
\end{equation}

This 5 order of magnitude broadband improvement in background phase noise represents approximately 2.5 orders of magnitude improvement in sensitivity to $g_{a\gamma\gamma}$. Taking the locked measurement of Fig.~\ref{Comparison} as an example, where the experimental parameters of the readout mode were $\beta_1=$ 0.83, $Q_1=$ 2550 and $P_{inc}=$ 10 dBm, we derive a minimum noise temperature of 360 K at the minimum value of $S_y$ with respect to the Fourier frequency ($\frac{\delta\omega_a}{2\pi}=~$3.1 kHz). With more effort in the design of the frequency feedback control loops and filters, it has been shown that this range could be extended \cite{Ivanov2009}. Without this extra effort, sensitivity would be lost unless minute tuning steps were made to localize datataking to this small window of low-noise Fourier space. This issue, combined with the difficulty of keeping the system locked as we tune, led us to conclude that in this mass range, which requires difference frequencies of 200 MHz or more, the power technique is a simpler method to implement, while achieving optimal sensitivity.

\section{General Calculation for Overlap Functions with $m=0$}
\label{appendix:B}
By solving the wave equation for a cylindrical cavity resonator one can find the unit vectors for the various resonant modes \cite{Kakazu96}. Analytical expressions for the unit vectors of the $TM_{0n0}$ modes are given by:
\begin{equation}
\mathbf{e}_{n_1}=\frac{J_0\left(\frac{\chi_{0n_1}}{r_c}r\right)}{J_1\left(\chi_{0n_1}\right)}\hat{z}~~\text{and}~~\mathbf{b}_{n_1}= \frac{J_1\left(\frac{\chi_{0n_1}}{r_c}r\right)}{J_1\left(\chi_{0n_1}\right)}\hat{\varphi};
\end{equation}
For $TM_{0,n,p}$ modes:
\begin{equation}
\begin{aligned}
&\mathbf{e}_{n_1p_1}=\frac{\sqrt{2}\pi p_1cJ_1\left(\frac{\chi_{0n_1}}{r_c}r\right)}{L\omega_1J_1\left(\chi_{0n_1}\right)}\sin \left(\frac{\pi  p_1 z}{L}\right)\hat{r} \\
&~~~~~~~~~~+\frac{\sqrt{2}\chi_{0n_1}cJ_0\left(\frac{\chi_{0n_1}}{r_c}r\right)}{r_c\omega_1J_1\left(\chi_{0n_1}\right)}\cos \left(\frac{\pi  p_1 z}{L}\right)\hat{z} \\
&\mathbf{b}_{n_1p_1}= \frac{\sqrt{2}J_1\left(\frac{\chi_{0n_1}}{r_c}r\right)}{J_1\left(\chi_{0n_1}\right)}\cos \left(\frac{\pi  p_1 z}{L}\right) \ \hat{\varphi};
\end{aligned}
\end{equation}
And for $TE_{0,n,p}$ modes:
\begin{equation}
\begin{aligned}
&\mathbf{e}_{n_0p_0}=-\frac{\sqrt{2}J_1\left(\frac{\chi_{0n_0}^\prime}{r_c}r\right)}{J_0\left(\chi_{0n_0}^\prime\right)}\sin \left(\frac{\pi  p_0 z}{L}\right)\hat{\varphi}\\
&\mathbf{b}_{n_0p_0}=\frac{\sqrt{2}cp_0\pi J_1\left(\frac{\chi_{0n_0}^\prime}{r_c}r\right)}{L\omega_0J_0\left(\chi_{0n_0}^\prime\right)}\cos \left(\frac{\pi  p_0 z}{L}\right) \ \hat{r} \\
&~~~~~~~~~~~-\frac{\sqrt{2}c\chi_{0n_0}^\prime J_0\left(\frac{\chi_{0n_0}^\prime}{r_c}r\right)}{r_c\omega_0J_0\left(\chi_{0n_0}^\prime\right)}\sin \left(\frac{\pi  p_0 z}{L}\right)\hat{z}.
\end{aligned}
\end{equation}
Here $J_{i}(x)$ is the $i^{th}$ Bessel function, $\chi_{a,b}$ represents the $b^{th}$ root of the $a^{th}$ Bessel function, $\chi_{a,b}'$ represents the $b^{th}$ root of the derivative of the $a^{th}$ Bessel function, $\omega$ is the resonant frequency, c is the speed of light, L is the cavity height and r is the cavity radius.

Thus, if the readout mode is a $TM_{0,n_1,0}$ mode and the background mode is a $TE_{0,n_0,p_0}$ mode, then the overlap functions are non-zero when $p_0$ is odd, and given by
\begin{equation}
\xi_{01}=\frac{1}{V}\int\mathbf{e}_{n_0p_0}\cdot\mathbf{b}_{n_1}dV=\frac{4\sqrt{2}\chi_{0n_0}^\prime}{p_0\pi(\chi_{0n_0}^{\prime 2}-\chi_{0n_1}^2)},
\label{Olap01}
\end{equation}
\begin{equation}
\xi_{10}=\frac{1}{V}\int\mathbf{e}_{n_1}\cdot\mathbf{b}_{n_0p_0}dV=\frac{4\sqrt{2}\chi_{0n_0}^\prime}{p_0\pi(\chi_{0n_0}^{\prime 2}-\chi_{0n_1}^2)}\frac{\omega_1}{\omega_0}.
\label{Olap10}
\end{equation}
In the more general case where the readout mode is a $TM_{0,n_1,p_1}$ mode ($p_1\ge 1$), and the background mode is again a $TE_{0,n_0,p_0}$ mode ($p_0\ge 1$), then the overlap functions are non-zero when the sum of $p_1+p_0$ is odd, and given by,
\begin{equation}
\xi_{01}=\frac{1}{V}\int\mathbf{e}_{n_0p_0}\cdot\mathbf{b}_{n_1p_1}dV=\frac{8p_0\chi_{0n_0}^\prime}{(p_0^2-p_1^2)\pi(\chi_{0n_0}^{\prime 2}-\chi_{0n_1}^2)},
\label{Olap01b}
\end{equation}
\begin{equation}
\xi_{10}=\frac{1}{V}\int\mathbf{e}_{n_11p_1}\cdot\mathbf{b}_{n_0p_0}dV=\frac{8p_0\chi_{0n_0}^\prime}{(p_0^2-p_1^2)\pi(\chi_{0n_0}^{\prime 2}-\chi_{0n_1}^2)}\frac{\omega_1}{\omega_0}.
\label{Olap10}
\end{equation}
In general $\xi_{10}=\xi_{01}\frac{\omega_1}{\omega_0}$, however for upconversion $\omega_1\approx\omega_0$ so $\xi_{10}\sim\xi_{01}$.

\section{Uncertainty Analysis}
\label{appendix:C}
We here give our best estimate of systematic uncertainty in the presented experiment, taking in many cases worst-case values for uncertainty of parameters. As per Eqn.~(\ref{AxionPSD}), $g_{a \gamma \gamma}$ may be expressed by

\begin{equation}
\begin{split}
&g_{a\gamma\gamma} \propto \frac{1}{G} \left(\sqrt{\frac{1+\beta_1}{\beta_1}}\right) \left( \frac{1+\beta_0}{\sqrt{\beta_0}} \right) \frac{1}{\sqrt{Q_{L_0}}} \frac{1}{\sqrt{Q_{L_1}}} \frac{1}{\xi} \sqrt{\mathcal{L}}\\
&\mathcal{L} = \frac{1}{1+4 Q_{L_1}^2\left(\frac{\delta f_a}{f_1}\right)^2}
\label{uncertaintyparameters}
\end{split}
\end{equation}

where $\mathcal{L}$ encodes filtering of the axion-induced power by the Lorentzian readout mode and $G$ is the gain between the readout port and the digitzer, which is used to retrieved $P_{cav}$. Terms with negligible uncertainty have been omitted.

The uncertainty on $\mathcal{L}$ was estimated by recognising that the extracted central readout mode frequency $f_1$, is subject to fitting error. $f_1$ (as it contributes to informing $\mathcal{L}$) is the greatest contributer to uncertainty as the trace from which it is extracted is produced from only 17 seconds of data, and is therefore fairly noisy. Taking the standard deviation of the ensemble of fitted $f_1$ values to be the error, we find the associated uncertainty in the position of the Lorentzian can produce underestimation of $P_{cav}$, especially in the region of full-width-half-maximum (FWHM), where $dP/dF$ is maximal. At the FWHM point, the Lorentzian centralization error can cause a 28\% underestimation in $P_{out}$. However, across the spectrum, the magnitude of error is on average 22\% which is the value we use in the uncertainty calculation. These fractional errors were found in the conservative case of the highest quality factor and highest central frequency. The true drift in $f_1$ over this measurement period produces relatively insignificant error compared to fitting error. In future searches, we may take longer averaging periods in order to reduce this fitting error.

The error in $\xi$, which is a parameter comparable to form factor (C), is taken from previous haloscope searches to be $\sim 5\%$ \cite{bartram2021search,Bartram2021,Quiskamp2022}. However, due to the relative geometric simplicity of our cavity, being an empty cylinder, compared to these haloscopes which feature internal tuning rods, we in fact expect our experimental uncertainty to be lower. In any case, it would be valuable to verify the resonant mode shapes via the bead pull method, which may be a topic of further work.

The errors in $\beta$ and $Q$ are conservatively estimated from previous haloscope experiments to be $\sim 10\%$ and $\sim 20\%$ respectively \cite{bartram2021search,Bartram2021,Quiskamp2022} and the error in the linear gain of the amplifier is taken to be $\sim 5\%$ which implies gain stability within 0.2 dB.
\begin{table}
\centering
\caption{Major Uncertainties on Parameters Contributing to $g_{a\gamma\gamma}$}
\label{tab:uncertainties}
\begin{tabular*}{\linewidth}{l@{\extracolsep{\fill}}ll}
\toprule
        & \text{Fractional Uncertainty}  \\ \midrule\midrule
$G$ 							  &   0.05      \\ 
$f(\beta_0)$  	       &      0.07        \\ 
$g(\beta_1)$ 		   &     0.16             \\ 
$\mathcal{L}$		   &   0.22               \\ 
$\xi$ 		   &   0.05         \\ 
$Q_{L}$ 		 &   0.2             \\  \bottomrule
\end{tabular*}

\end{table}

To simplify the uncertainty calculation, errors in the input coupling and output coupling expressions of Eqn.~(\ref{uncertaintyparameters}) in brackets (labelled $f(\beta_0)$ and $g(\beta_0)$ in Table~\ref{tab:uncertainties}) were estimated by finding the maximal fractional error on these expressions over all recorded $\beta$ when $\frac{\delta \beta}{\beta}$ is taken to be $0.1$. This maximum is conservatively taken to be the uncertainty on these expressions over all data-taking. The fractional uncertainties of the contributing parameters are summarized in Table~\ref{tab:uncertainties}.

Taking the quadrature average of these fractional uncertainties leads to our systematic uncertainty on $g_{a\gamma\gamma}$:

\begin{equation}
\begin{split}
&\frac{\delta g_{a\gamma\gamma}^{exc}}{g_{a\gamma\gamma}^{exc}} \approx \Biggl( \left(\frac{\delta G}{G} \right)^2 +\left(\frac{\delta f(\beta_0)}{f(\beta_0)} \right)^2\\
& + \left(\frac{\delta g(\beta_1)}{g(\beta_1)} \right)^2 + \left(\frac{1}{2} \frac{\delta \mathcal{L}}{\mathcal{L}} \right)^2 +\\
& \left(\frac{\delta \xi}{\xi} \right)^2 + \left(\frac{1}{2} \frac{\delta Q_{L_0}}{Q_{L_0}} \right)^2 + \left(\frac{1}{2} \frac{\delta Q_{L_1}}{Q_{L_1}} \right)^2 \Biggr)^{1/2} \approx 26 \%
\end{split}
\end{equation}
\vspace{10mm}
\bibliography{Bib.bib}
\bibliographystyle{unsrt}

\end{document}